\documentclass[twocolumn,showpacs,preprintnumbers,amsmath,amssymb]{revtex4}

\usepackage{amsmath,bbm,graphicx}
\usepackage{subfigure}

\newcommand{\bra}[1]{\langle #1|}
\newcommand{\ket}[1]{|#1\rangle}
\newcommand{\braket}[1]{\langle #1 \rangle}
\newcommand{\id}{\mathbbm{1}}
\newcommand{\bbe}{\mathbbm{e}}
\newcommand{\bbi}{\mathbbm{i}}
\newcommand{\tr}{\text{tr}}

\begin{document}
\title{Probing the Quantumness of Channels with Mixed States}
\date{\today}

\author{Hauke H\"aseler and Norbert L\"utkenhaus}
\affiliation{
 Institute for Quantum Computing and Department of Physics and Astronomy, University of Waterloo, Canada\\
Quantum Information Theory Group, University of Erlangen, Germany}

\begin{abstract}
We present an alternative approach to the derivation of benchmarks for quantum channels, such as memory or teleportation channels. Using the concept of effective entanglement and the verification thereof, a testing procedure is derived which demands very few experimental resources. The procedure is generalized by allowing for mixed test states. By constructing optimized measure and re-prepare channels, the benchmarks are found to be very tight in the considered experimental regimes.
\end{abstract}


\maketitle

\section{Introduction}
The field of quantum information science has brought forward a range of interesting information processing protocols, many of which have already been demonstrated experimentally in a multitude of physical implementations. In quantum communication, the most prominent examples are (arguably) quantum key distribution \cite{bennett84a,bennett92a,grosshans03a}, teleportation \cite{bennett93a,bouwmeester97a,braunstein98a}, and quantum memory \cite{julsgaard04a,appel08a} experiments, all of which were implemented in discrete (qubit) as well as continuous-variable settings.

All of these protocols are based on the fact that information is transmitted in the form of quantum states, either by directly sending non-orthogonal states or by using pre-shared entanglement. This stands in contrast to \emph{measure and re-prepare} channels, which convert incoming states to classical data by means of measurements and re-prepare new quantum states accordingly. This conversion to classical data destroys any advantage which quantum protocols have over classical ones.

This observation leads to a natural benchmark criterion: An experimental implementation of a quantum channel is successful only if it can outperform any measure and re-prepare channel. How to quantify the performance of a channel is not trivial. For memory or teleportation channels, this is typically done with the \emph{average fidelity} \cite{braunstein00a}. For a given ensemble of test states $\{ p_i, \rho_i^{in} \}$, and corresponding output states $\rho_i^{out}$, the average fidelity is defined as $\bar{F} = \sum_i p_i F(\rho_i^{in},\rho_i^{out})$.

For continuous-variable protocols, a maximization of the average fidelity for measure and re-prepare strategies was first performed by Hammerer \emph{et al.} for a Gaussian distribution of coherent test states \cite{hammerer05a}. Namiki \emph{et al.} recently derived an extension for channels of non-unity gain \cite{namiki08b}. Due to recent interest in the storage of squeezed light \cite{appel08a,honda08a}, fidelity-based benchmarks were derived for different ensembles of squeezed states \cite{namiki08a,adesso08a,calsamiglia09a,owari08a}. Clearly, finding the optimal measure and re-prepare channel is a challenging task even for input ensembles, which are very small or highly symmetric. Moreover, each benchmark is tied to an input ensemble in the sense that a change in the input ensemble requires a new optimization of the average fidelity. In addition, the fidelity can be hard to measure in an experiment, especially in the continuous-variable setting.

In this paper, we use a different technique to derive benchmarks which are particularly simple to implement experimentally. Our analysis mainly relies on two observations: Firstly, any measure and re-prepare scheme acts as an entanglement-breaking channel \cite{horodecki03a}. Therefore, it suffices to demonstrate that a memory or teleportation setup can preserve entanglement. Secondly, every source of non-orthogonal states admits a theoretical entanglement-based description, which is typically referred to as \emph{source replacement} or as \emph{effective entanglement} \cite{curty04a,grosshans04a,moroder06a}. This means that we can devise benchmarks based on the verification of entanglement without the need to generate actual entangled states in the laboratory.

A general benchmark criterion should accommodate mixed test states, to cover cases where the generation of pure test states is experimentally infeasible. For example, initially pure test states may be subject to a noisy environment before they reach the quantum channel to be tested. Another example of this is the generation of squeezed light, where losses in the preparation process lead to mixing. We therefore extend the technique of source replacement to sources which emit mixed states.

For the verification of effective entanglement, there exists a large number of methods, each with their particular advantages \cite{guhne09a}. Here, we will use the so-called expectation value matrix (EVM) criterion \cite{rigas06a, haseler08a}, which is tailored for situations where state tomography is not available. We show how to extend this criterion to incorporate sources of mixed states. The methods are exemplified with the help of three different physical settings: a qubit protocol, a source of displaced thermal states and a source of squeezed thermal states.

This paper is structured as follows. Section \ref{sec:pts} gives a review of the concept of effective entanglement for sources of pure states. Furthermore, the expectation value matrix method is reviewed and applied to a source of squeezed vacuum states, which may be used to benchmark quantum memory experiments. Section \ref{sec:mss} presents a formalism for source replacement and entanglement verification for mixed test states. Using the results from the pure test states, this formalism is developed in steps, which approximate the actual problem increasingly well. In Sec.~\ref{sec:oebc}, the effects of entanglement-breaking channels are considered. By optimizing the possible measurement outcomes induced by such channels, the whole domain compatible with separable states can be identified. This is a tool for assessing the strength of the developed entanglement criteria. Finally, Sec.~\ref{sec:d} contains a conclusion and a discussion of the results.

\section{Pure Test States}\label{sec:pts}
A test for quantum channels using pure test states was proposed in Ref.~\cite{haseler08a}. It relies on an entanglement-based source description and the subsequent verification of entanglement. In this section, we start with an outline of the proposed test procedure and a brief review of the entanglement verification method. We then investigate the application of the method to quantum memories for squeezed light.

The general setup is as follows: To probe the quantum channel, we employ a source of (non-orthogonal) test states $\ket{\psi_i^{in}}$, which are chosen at random from a fixed set $\{ \ket{\psi_i^{in}} \}_{i=0}^N$ of cardinality $N$. Each state is sent with a predetermined probability $p_i$. In quantum key distribution, this ensemble of test states is determined by the protocol, since it must be possible to exclude intercept-resend attacks using the same classical data that will ultimately be used to generate secret key. For quantum memories, the input ensemble can be chosen freely, in accordance with experimental feasibility. The key to the test procedure is the fact that every source of non-orthogonal test states admits an entanglement-based description. In this thought setup, an entangled source state $\ket{\psi^{src}} = \sum_{i=0}^N \sqrt{p_i} \ket{i}_A\ket{\psi_i^{in}}_B$ is generated, and we label the two subsystems by A (Alice) and B (Bob). The correct test state ensemble can effectively be prepared by measuring system $A$ in the basis $\{\ket{i}_A\}$. This \emph{effective entanglement} presents a natural test for a quantum channel, which outputs states $\rho_i^{out}$ for each input $\ket{\psi_i}$ from the test state ensemble. If we can verify entanglement between system $A$ and the output states, the quantum channel did not act as a measure and re-prepare channel.

\subsection{Review of the expectation value matrix method}
The verification of entanglement between Alice and Bob bears two major difficulties. Firstly, the dimensions of $\mathcal{H}_A$ and $\mathcal{H}_B$ are, in general, different. The dimension of system $A$ is at most equal to the number $N$ of pure test states. The dimension of the output system $B$ may be much higher. The second difficulty arises from partial information. In most situations, it is challenging to perform state tomography on the output states, especially in optical implementations (see, e.g., \cite{leonhardt97a}). If the measurements performed by Alice and Bob only supply partial knowledge, no unique state can be assigned to a set of measurement data. Instead, there will be a set of compatible states (an equivalence class). The task of verifying entanglement is now mapped to checking whether this equivalence class contains a separable state.

The (EVM) \cite{rigas06a} is a tool to perform exactly this task and it works for arbitrary dimensions and for a broad class of observables. Its construction depends on the test state ensemble and on Bob's measurement operators. That is, the entries of the EVM are defined, as a function of the effective bipartite state $\rho_{AB}$, by
\begin{equation}\label{eq:evm}
	[\chi(\rho_{AB})]_{ijkl} = \text{Tr}(\rho_{AB} \hat{A}_i^\dagger \hat{A}_k \otimes \hat{B}_j^\dagger \hat{B}_l).
\end{equation}
The operator sets $\{ \hat{A}_i \}$ and $\{ \hat{B}_j \}$ must be chosen according to the measurement operators given by the protocol. For a binary input ensemble, the most compact set $\{ \hat{A}_i \}$ is $\{ \ket{\phi}\bra{0}, \ket{\phi}\bra{1} \}$, where $\ket{\phi}$ is a generic qubit state, which does not actually come into play due to the structure in Eq.~(\ref{eq:evm}). For each possible set of measurement outcomes, the EVM is designed to be a compact representation of the corresponding equivalence class of states. The membership of separable states in this equivalence class is then tested by using the EVM in conjunction with a positive, but not completely positive map \cite{horodecki96a}, such as the partial transposition map \cite{peres96a}. By construction \cite{haseler08a},
\begin{equation}
	\chi(\rho_{AB}) \ge 0 \quad \forall \ \rho_{AB}.
\end{equation}
Therefore, violation of the condition
\begin{equation}\label{eq:chinpt}
	\chi(\rho_{AB}^{T_A}) \ge 0
\end{equation}
is a sufficient condition for entanglement in $\rho_{AB}$. To check this condition, it must be possible to relate $\chi(\rho_{AB}^{T_A})$ to the measurement outcomes stored in  $\chi(\rho_{AB})$. In the above choice of operators for Alice, $\chi(\rho_{AB}^{T_A})$ is related to $\chi(\rho_{AB})$ by a simple block transposition.

If not all entries of $\chi(\rho_{AB})$ are experimentally accessible, the condition (\ref{eq:chinpt}) can be checked with the help of semidefinite programming \cite{vandenberghe96a}.

\subsection{Pure squeezed test states}

One application of the EVM method was presented in Refs.~\cite{haseler08a} and \cite{rigas06a}, where the two coherent states $\ket{\alpha}$ and $\ket{-\alpha}$ are used to test a channel for quantum cryptography against intercept-resend attacks. Here, we investigate the similar task of testing a quantum memory with squeezed states, for which the same method can be employed. A convenient choice of test-states are the squeezed vacuum and the phase-rotated squeezed vacuum,
\begin{align}
	\ket{\psi_0^{in}} & = \hat{S}(r^{in}) \ket{0}\\
	\ket{\psi_1^{in}} & = \hat{S}(-r^{in}) \ket{0}.
\end{align}
If two orthogonal quadratures $\hat x$ and $\hat p$ of the output light are measured, and the first and second moments of those operators are recorded, an EVM can be constructed analogously to Ref.~\cite{haseler08a}.

It is at this point instructive to find a parameterization of typical measurement outcomes. We then show for which parameter regimes entanglement verification is possible. Since the operators $\hat{x}, \hat{p}, \hat{x}^2$, and $\hat{p}^2$ are measured on two possible output states, the measurement outcomes are characterized by eight parameters. However, if the quantum device at hand does not introduce displacements, all first moments will be zero. Moreover, typical quantum channels are oblivious of the phase space orientation of the input states, so the output variances will be related by
\begin{align}\label{eq:symm1}
	{\rm Var}_{\rho_0^{out}}(\hat x) & = {\rm Var}_{\rho_1^{out}}(\hat p) \\ \label{eq:symm2}
	{\rm Var}_{\rho_0^{out}}(\hat p) & = {\rm Var}_{\rho_1^{out}}(\hat x).
\end{align}
The EVM then takes the simple form
\begin{equation}\label{eq:chi}
	\begin{pmatrix}
		{\rm Var}_0(\hat x) & b_1 + \frac{\bbi}{2} & c_1 & c_2 + \frac{\bbi}{2} s \\
		b_1 - \frac{\bbi}{2} & {\rm Var}_0(\hat p) & c_2 - \frac{\bbi}{2} s & c_3 \\
		c_1^* & c_2^* + \frac{\bbi}{2} s & {\rm Var}_0(\hat p) & b_2 + \frac{\bbi}{2} \\
		c_2^* - \frac{\bbi}{2} s & c_3^* & b_2 - \frac{\bbi}{2} & {\rm Var}_0(\hat x)
	\end{pmatrix},
\end{equation}
where $s$ denotes the overlap $\braket{-r^{in}|r^{in}}$ of the test states. The operators $\hat{B}_j$ are chosen from the set $\{\hat{x}, \hat{p}\}$, since the first moments need not be included. The parameters $b_i$ $\in \mathbbm{R}$ and $c_i$ $\in \mathbbm{C}$ are not experimentally accessible, so that the condition (\ref{eq:chinpt}) must be checked by numerical evaluation of the resulting semidefinite program.

\begin{figure}
	\centering
		\includegraphics[width=0.42\textwidth]{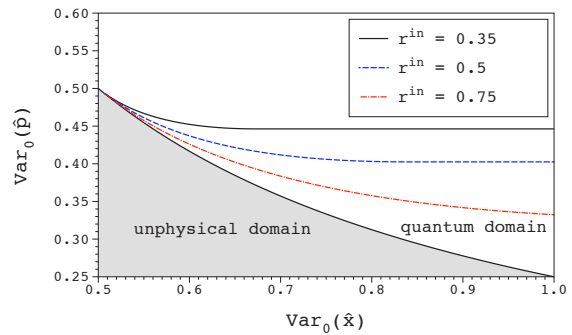}
	\caption{(Color online) Possible measured variances from two squeezed test states. For higher degrees of input squeezing, higher noise reductions must be observed to verify effective entanglement. The shaded area is excluded by the uncertainty principle.}
	\label{fig:pure}
\end{figure}

Now, two parameters, say ${\rm Var}_{\rho_0^{out}}(\hat x)$ and ${\rm Var}_{\rho_0^{out}}(\hat p)$, suffice to specify a distinct set of measurement outcomes. Figure \ref{fig:pure} shows the area of physically allowed measurement data in this parameterization. The gray area is excluded by the Heisenberg uncertainty relation. Using the EVM method, we can identify parameter pairs which correspond to effective entangled states. These parameter pairs form the \emph{quantum domain}, in which the quantum channel indeed operates in a non-classical manner. The boundary of the quantum domain shifts as the test-state-overlap $s$ is varied.

The plots show that the observation of squeezing alone, i.e., ${\rm Var}_{\rho_0^{out}}(\hat p) < 1/2$, does not suffice. For higher degrees of squeezing in the test states, one must also observe more squeezing in the output states to verify operation in the quantum domain.\\

\subsection{Optimal measure and re-prepare channels}
The remainder of this section concerns those parameter pairs outside the quantum domain. Since the EVM method and indeed partial transposition form in general sufficient criteria for entanglement, but not necessary ones, points outside the quantum domain can fall into two categories: either they stem from effective entangled states which were not detected, or they are compatible with a classical channel, i.e., a measure and re-prepare strategy. In the above parameter space, the latter category leads to a \emph{classical domain}. By maximizing this classical domain, we can assess the strength of our test procedure.

Mathematically, a measure and re-prepare strategy is equivalent to an entanglement-breaking channel \cite{horodecki03a},
\begin{equation}\label{eq:ebchannel}
	\rho_k^{out} = \sum_i {\rm tr}(\rho_k^{in} \pi_i) \ket{\tilde{\psi}_i}\bra{\tilde{\psi}_i},
\end{equation}
where the operators $\pi_i$ form a positive operator valued measure (POVM), i.e., they are positive semidefinite and sum to the identity. Maximizing the classical domain now corresponds to the optimization of a suitable figure of merit by varying both the measurement operators $\{\pi_i\}$ and the re-prepared states $\{\ket{\tilde{\psi}_i}\}$. For the aforementioned case of coherent test states $\ket{\alpha}$ and $\ket{-\alpha}$, the figure of merit is the excess noise, or variance broadening, measured by Bob and the optimal entanglement-breaking channel minimizes this excess noise. This optimal channel was found in Ref.~\cite{haseler08a} to be comprised of a minimum error discrimination POVM \cite{helstrom76a} and the resending of displaced squeezed states. For the case of two squeezed test states, the optimal entanglement-breaking channel is the one which minimizes ${\rm Var}_{\rho_0^{out}} (\hat p)$ for each fixed value of ${\rm Var}_{\rho_0^{out}} (\hat x)$ (c.f.~Fig.~\ref{fig:pure}).

\begin{figure}
	\centering
		\includegraphics[width=.42\textwidth]{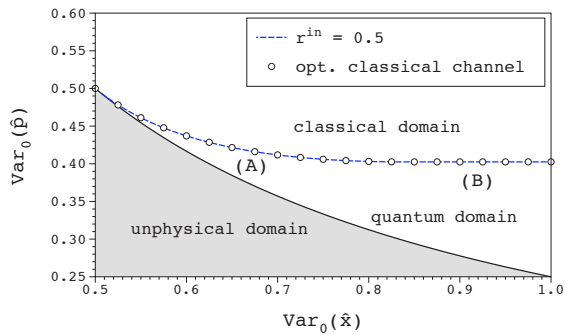}
	\caption{(Color online) Optimal measure and re-prepare strategy for two squeezed test states. The boundaries of the classical and the quantum domains coincide. Optimal measurements in region (A): minimum error discrimination; region (B): unambiguous state discrimination}
	\label{fig:sqattack}
\end{figure}

Let us first consider the re-prepared states $\{\ket{\tilde{\psi}_i}\}$. Since the figure of merit is a variance, the re-prepared states must be minimum-uncertainty states. Furthermore, since Bob expects all first moments to be zero, displacements in $\{\ket{\tilde{\psi}_i}\}$ have the effect of increasing the quadrature variances (see, e.g., \cite{barnett97b}). This leaves the class of squeezed vacuum states, with the squeezing axis aligned with the axis of Bob's quadrature detection. Therefore, each re-prepared state $\ket{\tilde{\psi}_i}$ can be parameterized by a single real parameter, namely, its squeezing parameter $r_i$. Let us now turn to the POVM $\{ \pi_i \}$. Since there are only two pure test states, each POVM element is described by a two-dimensional matrix. Assuming that the number of POVM elements is small, the optimal solution can easily be found numerically. Explicitly, the figure of merit is given by
\begin{equation}
	{\rm Var}_{\rho_0^{out}} (\hat p) = \frac{1}{2}\sum_i {\rm tr} (\pi_i \rho_0^{in}) \exp (-2 r_i),
\end{equation}
which we wish to minimize for each fixed value of ${\rm Var}_{\rho_0^{out}} (\hat x)$ under the constraints
\begin{align}
	\pi_i & \ge 0, \\
	\sum_i \pi_i & = \id.
\end{align}
The results of a numerical optimization are shown in Fig.~\ref{fig:sqattack}, with the number of POVM elements set to four. We see that this optimized measure and re-prepare strategy reaches the boundary of the quantum domain. Therefore, any experimental outcome in our parameterization is unambiguously sorted into one of three domains: the unphysical domain, the quantum domain as determined by the EVM method, and the classical domain.

It is natural to ask whether this optimal POVM corresponds to minimum error discrimination, in analogy to the coherent-state-protocol. It turns out that this is true only partially. The boundary of the quantum domain shown in Fig.~\ref{fig:sqattack} is a concatenation of a curved part (A) and a straight line with slope zero (B). This behavior arises from the fact that the first eigenvalue of the EVM to become negative is different in regions (A) and (B). Therefore, the optimal measure and re-prepare strategies and in particular the POVMs are also likely to be different in both regions. It turns out that the optimal measurement for region (A) is indeed the minimum error discrimination. For the measurement in region (B), we recall that the measured variances are linear functions of $\rho_0^{out}$, so that any convex combination of two measure and re-prepare strategies results in another valid entanglement-breaking channel. Therefore, finding one particular strategy for which ${\rm Var}_{\rho_0^{out}} (\hat x)$ diverges, while ${\rm Var}_{\rho_0^{out}} (\hat p)$ remains finite, explains the straight line with slope (approaching) zero. Such a strategy involves the unambiguous discrimination of the two input states \cite{jaeger95a}. When the measurement outcome is inconclusive, vacuum states are re-prepared, and whenever $\rho_0^{in}$ is identified unambiguously, an infinitely squeezed state is re-prepared. This way, ${\rm Var}_{\rho_0^{out}} (\hat x)$ tends to infinity while ${\rm Var}_{\rho_0^{out}} (\hat p)$ remains finite.\\

With the combination of minimum error measurement and unambiguous state discrimination, we can show analytically that the EVM method witnesses all entangled states which are detectable by the given measurements. The resulting quantum domain cannot be enlarged by any other verification method. This is in accordance with the results for two coherent test states $\ket{\alpha}$ and $\ket{-\alpha}$.

\section{Mixed State Sources}\label{sec:mss}
In this section, we consider sources of mixed test states. In a communication context, this mixing operation may be useful and simplify the transmission of secret messages \cite{renner05b}. More common, however, are situations where it may be desirable to use pure states, but not experimentally feasible.

Any source of mixed quantum states $\rho_i^{in}$ can be thought of as a source of pure states $\ket{\psi_i^{in}}$, followed by a mixing operation. This picture shows a natural way to include mixed-state sources in the EVM-method. For the entanglement-based description and the construction of the EVM, we can consider the pure states $\ket{\psi_i^{in}}$, and attribute the mixing process to the quantum channel to be tested. In this way, the method from the previous section can be used. However, the mixing process will introduce an additional degradation of the channel. Therefore, any quantum benchmark derived in this way will be valid, but not tight, since it actually is a benchmark on the concatenation of the mixing process and the action of the quantum channel.

\subsection{Mixed test states and purifications}
What is a better way to describe the mixed-state sources? Since the EVM-method works very well for sources of pure states, we would like to build on the language developed in the previous section. We start by fixing some notations. Let the test state ensemble be given by the probabilities $p_i$ and the corresponding test states $\rho_i^{in}$. We can decompose each test state into pure states as $\rho_i^{in} = \sum_j q_j^{(i)} \ket{j^{(i)}}\bra{j^{(i)}}$.

\begin{figure}
	\centering
	\subfigure[mixture of pure states]{\includegraphics[height=1.8cm]{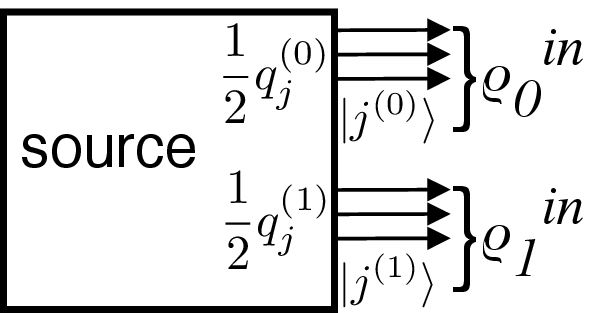}
	\label{fig:msource1}}
	\subfigure[part of a purification]{\includegraphics[height=1.8cm]{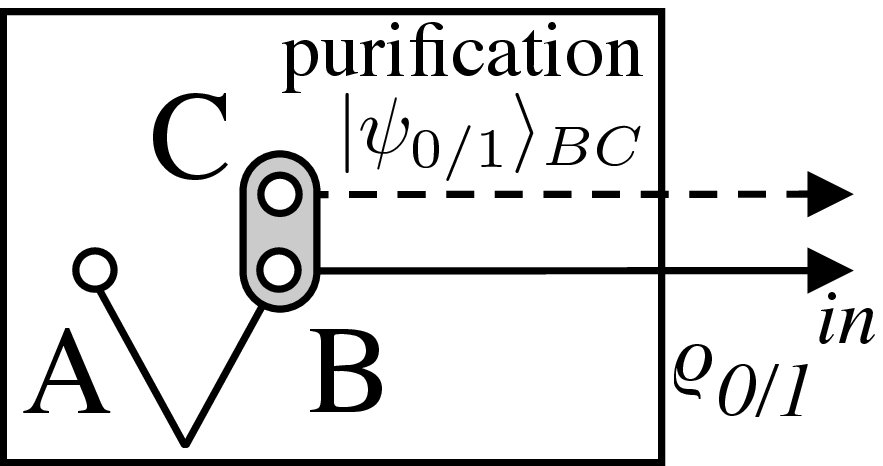}
	\label{fig:msource2}}
	\label{fig:sources}
	\caption{Each mixed state emitted from the source can be regarded as a mixture of pure states, or as part of a higher-dimensional pure state (purification)}
\end{figure}

Considering each mixed state emitted from the source as a statistical mixture of pure states, we see that the source actually emits pure states from the set $\{ \ket{j^{(i)}} \}$, with probabilities $\{p_i\ q_j^{(i)}\}$ [see Fig.~\ref{fig:msource1}]. An entanglement-based description of the source is now straight-forward. Problems arise when it comes to the construction of the EVM, since Bob's measurements cannot resolve which pure state in the decomposition of a test state was sent. In other words, the measurement outcomes may only be conditioned on the index $i$, but not on the index $j$. This makes the direct construction of an EVM impossible.

There is, however, an alternative way to connect mixed states and pure states. Every mixed state $\rho$ can be thought of as a part of a higher-dimensional pure state $\ket{\psi}$, which is called a \emph{purification} of $\rho$ \cite{nielsen00a} . Hence, a possible simplified description of the actual mixed-state source is a source which emits purifications of the test states. This is illustrated in Fig.~\ref{fig:msource2}. The test-state source, which emits mixed states on system $B$ is approximated by a source  which emits purifications $\ket{\psi_i}_{BC}$, where the dimension of system $C$ is at most equal to the dimension of system $B$. For simplicity, we will again concentrate on the smallest non-trivial input ensemble, which contains only two test states, each occurring with probability $1/2$. In an entanglement-based description, the resulting effective entangled state is
\begin{equation}\label{eq:puri}
	\ket{\psi^{src}} = \frac{1}{\sqrt{2}} (\ket{0}_A \ket{\psi_0}_{BC} + \ket{1}_A \ket{\psi_1}_{BC}),
\end{equation}
where the $\ket{\psi_i}_{BC}$ are purifications of the actual test states $\rho_i^{in}$, $i \in \{0,1\}$.

An EVM can now be constructed in direct analogy to the pure-state-case of the previous section. The fact that Bob does not have access to the auxiliary system is easily incorporated by appending $\id_C$ to each operator which enters the EVM. In other words, in the general form of a three-party EVM
\begin{equation}\label{eq:3evm}
	[\chi(\rho_{ABC})]_{ijklmn} = \tr(\rho_{ABC} \hat{A}_i^\dagger \hat{A}_l \otimes \hat{B}_j^\dagger \hat{B}_m \otimes \hat{C}_k^\dagger \hat{C}_n),
\end{equation}
the set $\{ \hat{C}_k \}$ has only one member, namely $\id_C$. This way, the auxiliary system and the use of purifications do not increase the size of the EVM when we compare it to the pure-state-case. The dependence of the EVM on the test-state-overlap $s$ is changed to a dependence on the overlap of the two purifications $\ket{\psi_0}_{BC}$ and $\ket{\psi_1}_{BC}$.

\subsection{Optimal purifications}
It should be noted that a purification is not unique, i.e., there is a freedom in the choice of $\ket{\psi_i}_{BC}$. It turns out that the entanglement criterion works best if $\ket{\psi_0}_{BC}$ and $\ket{\psi_1}_{BC}$ are chosen such that their overlap takes its maximum value. From Uhlmann's theorem \cite{uhlmann76a}, we know that
\begin{equation}
	\max_{\ket{\psi_0},\ket{\psi_1}} |\braket{\psi_0 | \psi_1}| = F(\rho_0^{in}, \rho_1^{in}),
\end{equation}
where the maximization runs over all purifications $\ket{\psi_i}$ of the states $\rho_i^{in}$, and $F$ denotes the fidelity
\begin{equation}\label{eq:fid}
	F(\rho_0, \rho_1) = \tr \sqrt{\sqrt{\rho_0} \rho_1 \sqrt{\rho_0}}.
\end{equation}

As mentioned above, a source which emits the purifications $\ket{\psi_i}_{BC}$ is an approximation to the true test-state source. This approximation leads to weakened benchmarks, since the two purifications can be distinguished more easily than the mixed test states, which facilitates measure and re-prepare strategies. Ideally, only system $B$ is emitted and the purifying system $C$ is retained in the source [see Fig.~\ref{fig:msource2}]. In other words, Alice has full access to the reduced density matrix $\rho_{AC}$. However, the above construction of the EVM does not contain the full matrix $\rho_{AC}$, but only $\rho_A$.

\subsection{Further improvements}
In order to witness all of the detectable entanglement, the information stored in the purifying system must be included in $\chi(\rho_{ABC})$. A natural way to achieve this is through measurements on system $C$. The corresponding measurement operators $\{\hat{C}_k\}$ can then be included in the EVM, c.f.~Eq.~(\ref{eq:3evm}). If a tomographically complete set $\{\hat{C}_k\}$ is chosen, $\rho_{AC}$ will be fully contained in $\chi(\rho_{ABC})$. However, care must be taken with this approach. The test states must not be conditioned on measurement outcomes of the system $C$. This would lead to a different test state ensemble. Also, expectation values on system $C$ are only available after tracing over Bob's system, since they reflect the knowledge of the reduced density matrix $\rho_{AC}$.

Below, we describe three physical setups: One qubit protocol, one setup with squeezed thermal states as test states and one setup with displaced thermal states.\\

\begin{figure}
	\centering
		\includegraphics[width=.3\textwidth]{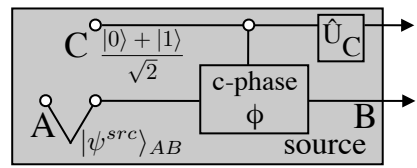}
	\caption{Source-replacement scheme for a source of mixed qubit states.}
	\label{fig:qubitsetup}
\end{figure}

\emph{Qubit states.} We begin by considering an example with qubit test states, which will serve to set the notations and to demonstrate the use of purifications. The setup is shown in Fig.~\ref{fig:qubitsetup}. Inside the test-state source, an entangled state $\ket{\psi^{src}}_{AB}$ is created, such that two pure qubit states $\ket{\psi_0^{in}}_B$ and $\ket{\psi_1^{in}}_B$ can be prepared by projective measurements on system $A$. It is convenient to choose them
\begin{align}
	\ket{\psi_0^{in}}_B & = \cos(\frac{\theta}{2}) \ket{0}_B + \sin(\frac{\theta}{2}) \ket{1}_B, \\
	\ket{\psi_1^{in}}_B & = \cos(\frac{\theta}{2}) \ket{0}_B - \sin(\frac{\theta}{2}) \ket{1}_B.
\end{align}

These states are mixed by the interaction with an auxiliary qubit system $C$ through a conditional phase gate. This auxiliary system is initially in the superposition $(\ket{0} + \ket{1})/\sqrt{2}$. By adjusting the phase $\phi$, the mixedness of the conditional states in system $B$ can be varied continuously. 

\begin{figure}
	\centering
		\includegraphics[width=.42\textwidth]{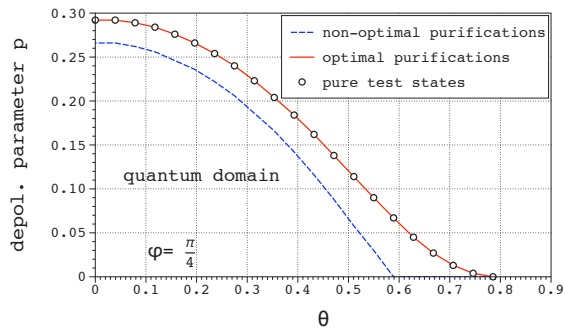}
	\caption{(Color online) Results of the entanglement verification for qubit states. The verifiable quantum domain for mixed states shrinks (dashed line) unless the optimal purifications are used (circles)}
	\label{fig:qubit}
\end{figure}

In this setting, $\phi = 0$ corresponds to pure test states. For this case, no purifying system is required. The results of the entanglement verification are shown in Fig.~\ref{fig:qubit}. Here, Bob can measure all three Pauli operators and his observations are modeled by a depolarizing channel,
\begin{equation}
	\rho_i^{out} = \Lambda^p (\rho_i^{in}) = p \frac{\id_B}{2} + (1-p) \rho_i^{in}.
\end{equation}
The figure shows the maximum value of the depolarizing parameter $p$ for which effective entanglement is still detected, for different values of the angle $\theta$. We observe that more channel noise can be tolerated when the signal states overlap more, similar to the observations for two phase-conjugated coherent states.

What happens when the test states become mixed, i.e., the conditional phase gate is in place with $\phi \not = 0$? The joint states $\ket{\psi_{i}}_{BC}$ of systems $B$ and $C$ after this gate will not be of product form anymore, and they are purifications of the mixed test states. The dashed line in Fig.~\ref{fig:qubit} shows that the verifiable quantum domain shrinks. However, the unitary freedom in the choice of the purifications has not been exploited yet. If we choose those two purifications with the maximum overlap, more effective entangled states are detected (circles in Fig.~\ref{fig:qubit}). In fact, the detected quantum domain is the same as for pure test states.

Finally, we can include the information contained in the purifying system $C$, by performing measurements on it. The Pauli operators are an obvious choice of measurement operators, since, together with the identity, they form an operator basis for $\mathcal{H}_C$. In practice, this amounts to including the set $\{\hat{C}_k \} = \{ \id_C, \sigma_x, \sigma_y \}$ in the EVM. It turns out that this larger EVM cannot detect any additional states, or in other words, the approximation with purifications is an equally strong criterion.

This a particularity of this example with qubits, which will be examined in Sec.~\ref{sec:oebc}. The next two examples show that the inclusion of the information stored in $\rho_{AC}$ can indeed improve the EVM-method. \\

\emph{Squeezed thermal states.} Squeezed light is of fundamental interest due to its non-classical character, which shows for example in sub-Poissonian photon number distributions and in uncertainties below the level of vacuum fluctuations. The generation of quadrature squeezed light by processes such as optical parametric amplification has become a standard technique. The use of two squeezed vacuum states to probe a quantum channel was described in the previous section. An extension of the method to mixed states is essential, since squeezed states become thermalized under the influence of losses. The resulting test states are
\begin{align}
	\rho_0^{in} & = \frac{1}{\bar{n} + 1} \sum_{n=0}^\infty \left( \frac{\bar{n}}{\bar{n} + 1} \right)^n \hat{S}(r^{in}) \ket{n} \bra{n} \hat{S}^\dagger(r^{in}) \\
	\rho_1^{in} & = \frac{1}{\bar{n} + 1} \sum_{n=0}^\infty \left( \frac{\bar{n}}{\bar{n} + 1} \right)^n \hat{S}^\dagger(r^{in}) \ket{n} \bra{n} \hat{S}(r^{in}),
\end{align}
i.e., they are characterized by their mean photon number $\bar n$ and the squeezing parameter $r^{in}$ (see, e.g., \cite{barnett97b}). The source replacement and the entanglement verification work very similarly to the above example with qubit states. The main differences are the following: While the qubit test states act on a two-dimensional Hilbert space, independently of the degree of mixing, two squeezed thermal states act on an infinite-dimensional Hilbert space, and each test state has full support. Furthermore, full tomography on the output states would result in an infinite-dimensional EVM, so, in analogy with the previous section, we consider detection of the first and second moments of the quadratures $\hat x$ and $\hat p$. For the typical measurement outcomes mentioned above, i.e., both quadrature operators have zero expectation values and the variances are related through Eqs. (\ref{eq:symm1}) and (\ref{eq:symm2}), the EVM for mixed states is of the same form as for pure states [Eq.~(\ref{eq:chi})]. Here, the operator set $\{ \hat{C}_k \}$ in the general form (\ref{eq:3evm}) has only one member, namely, $\id_C$.

\begin{figure}
	\centering
		\includegraphics[width=.42\textwidth]{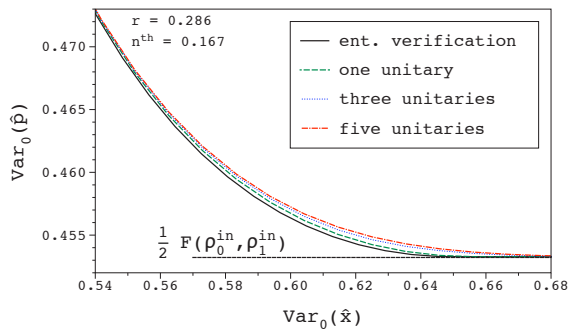}
	\caption{(Color online) Entanglement verification from different EVMs. Solid line: no information on the purifying system; dashed line: one generalized spin operator on the purifying system; dotted line: three generalized spin operators; dashed-dotted line: five generalized spin operators}
	\label{fig:unitaries}
\end{figure}

Results of the entanglement verification are shown in Fig.~\ref{fig:unitaries}. These results strongly resemble those from the previous section, where pure squeezed test states were considered. Indeed, the only difference is that the overlap of the pure test states is replaced by the fidelity of the mixed test states. We therefore note that the EVM-method can be extended to include sources of mixed states without introducing additional complexity. But, as mentioned above, we are merely approximating the source of two mixed test states by a source of two purifications of the test states.

We now proceed to include more of the information stored in $\rho_{AC}$ in the EVM. This is done by enlarging the set of operators $\{ \hat{C}_k \}$ which enter the construction of $\chi(\rho_{ABC})$. For the numerical implementation, it is convenient to choose unitary operators, and ideally, the set becomes large enough to form an operator basis for the space of operators on $\mathcal{H}_C$. However, since the dimension of this space is infinite, we are forced to restrict ourselves to a small subset. It is also unknown which operators to choose in order to achieve the biggest improvement of the entanglement criterion. We therefore choose the operators from a generic set, namely the generalized spin operators (see, e.g., \cite{pittenger04a}), which can be regarded as a generalization of the Pauli operators to higher dimensions. Each additional member in the set $\{ C_k \}$ enlarges the dimension of the EVM, and we have evaluated the entanglement criterion with up to five additional unitary operators on system $C$. Here, we worked in a truncated Fock basis with maximum photon number $n_{max} = 16$. The corresponding results are shown in Fig.~\ref{fig:unitaries}. It can be seen that the addition of operators on the purifying system $C$ does indeed improve the criterion. The more operators are included, the more entangled states are detected, but the size of $\chi$ grows and with it the number of free parameters. Therefore, there is a trade-off between the strength of the entanglement criterion and its complexity. It will, however, become clear in the next section that the curves shown in Fig.~\ref{fig:unitaries} are already close to the border of the classical domain.\\

\emph{Displaced thermal states.} Finally, we briefly examine the case of two displaced thermal states as test states. Experimentally, coherent states can be generated essentially noiselessly. Still, the addition of noise to the test states may be of advantage to the protocol at hand. Moreover, this second example in the continuous-variable domain will supplement the results for squeezed thermal states and allow us to draw a comparison.

In terms of Fock states, the two test states are given by
\begin{align}
	\rho_0^{in} & = \frac{1}{\bar{n} + 1} \sum_{n=0}^\infty \left( \frac{\bar{n}}{\bar{n} + 1} \right)^n \hat{D}(\alpha) \ket{n} \bra{n} \hat{D}^\dagger(\alpha) \\
	\rho_1^{in} & = \frac{1}{\bar{n} + 1} \sum_{n=0}^\infty \left( \frac{\bar{n}}{\bar{n} + 1} \right)^n \hat{D}(-\alpha) \ket{n} \bra{n} \hat{D}^\dagger(-\alpha).
\end{align}

\begin{figure}
	\centering
	\subfigure[pure vs. mixed test states]{\includegraphics[width=0.42\textwidth]{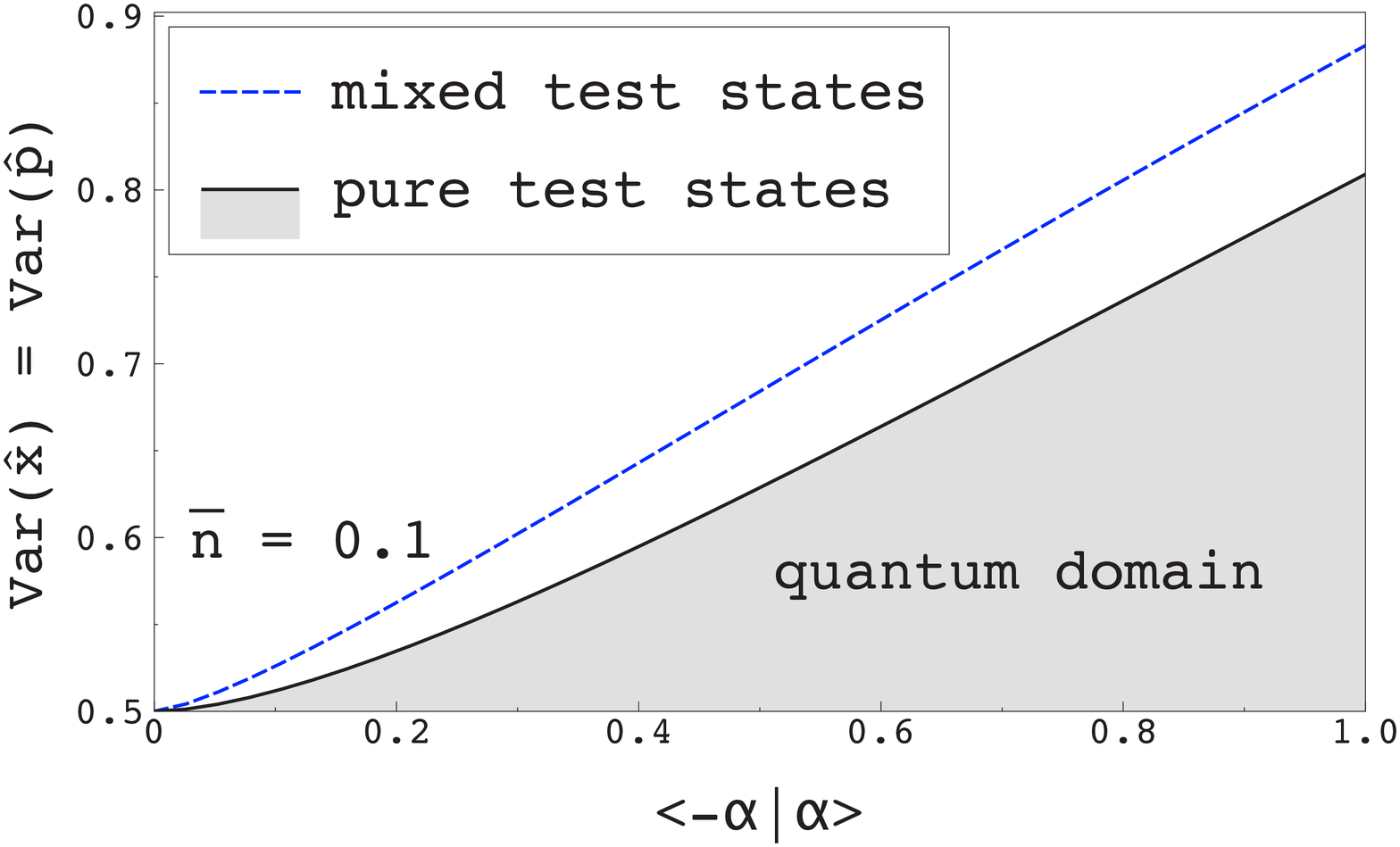}
	\label{fig:disp1}}
	\subfigure[quantum and classical domains]{\includegraphics[width=0.42\textwidth]{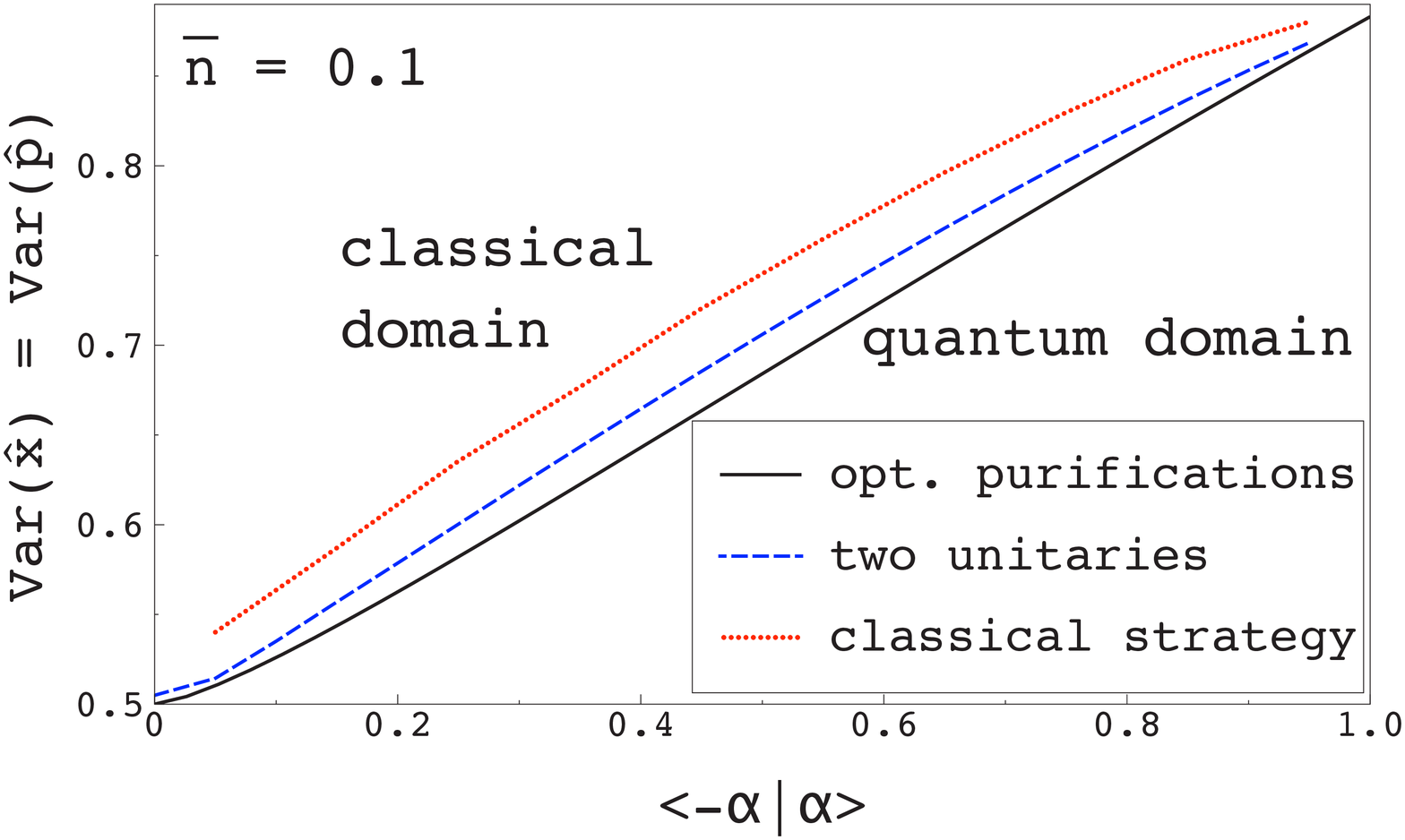}
	\label{fig:disp2}}
	\label{fig:disp}
	\caption{(Color online) Quantum domain for displaced thermal test states. \subref{fig:disp1}: the quantum domain increases in size for mixed states. \subref{fig:disp2}: improvements of the entanglement verification by adding two unitary operators to the EVM (dashed) and optimized measure and re-prepare strategy (dashed-dotted)}
\end{figure}

Here, $\hat{D}(\alpha)$ denotes the displacement operator \cite{barnett97b} and we take $\alpha$ to be real. These two test states pass through the quantum channel and, as before, homodyne detection is performed on the output states. We will, however, choose a different parameterization of the possible measurement outcomes in order to be consistent with \cite{rigas06a,haseler08a}. Specifically, the observed variances of the $\hat x$ and $\hat p$ quadratures are typically equal, and losses act as a simple down-scaling of the observed first moments. The results for pure test states \cite{rigas06a,haseler08a} are displayed in Fig.~\ref{fig:disp1}.

As before, a good approximation to the source of mixed test states is a source of corresponding purifications. Results of the entanglement verification using the optimal purifications are shown in Fig.~\ref{fig:disp1}. Remarkably, the verifiable quantum domain is enlarged compared to the setting with pure states. To be precise, two coherent test states $\ket{\alpha}$ and $\ket{-\alpha}$ lead to a smaller quantum domain than the same coherent states with added noise.

The strength of the entanglement criterion can be improved by including information on the purifying system $C$. As for the above example with squeezed thermal states, we enlarge $\chi(\rho_{ABC})$ with unitary operators $\hat{U}_C$ from the set of generalized spin operators. The dashed line in Fig.~\ref{fig:disp2} shows the results for two added unitary operators. Evidently, the quantum domain is significantly enlarged. But how close are these results to the boundary of the classical region, i.e., to those measurement outcomes compatible with a measure and re-prepare channel? The results of such a channel are indicated by the dashed-dotted line in Fig.~\ref{fig:disp2}. Here, the re-prepared states are displaced squeezed vacuum state, and the measurements are optimized numerically. A more detailed discussion of optimized measure and re-prepare channels is given in the next section.

We note that the results for displaced thermal states and squeezed thermal states are qualitatively very similar. The approximation of the true test-state source by a source of purifications leads to simple and reasonably strong entanglement criteria. In both cases, the addition of a small number of unitary operators which act on system $C$ pushes the boundary of the detected quantum domain close to the boundary of the classical domain.

\section{Optimized Entanglement-Breaking Channels}\label{sec:oebc}
As before for the pure test states, it is of interest to find the border of the quantum domain the converse way, i.e., by optimizing corresponding entanglement-breaking channels. According to Eq.~(\ref{eq:ebchannel}), an entanglement-breaking channel is characterized by the POVM $\{\pi_i\}$ and the re-prepared states $\{\ket{\tilde{\psi}_i}\}$. In this section, we will derive optimized measure and re-prepare strategies for the qubit setup and the squeezed state setup described above.

For pure qubit test states, it is again the minimum error POVM which leads to the optimal strategy. If the re-prepared states $\{\ket{\tilde{\psi}_i}\}$ are chosen of the same form as the original test states, the noise added by the measurement errors can be expressed as the action of a depolarizing channel. This allows a comparison between the boundaries of the quantum domain as shown in Fig.~\ref{fig:qubit} and the classical domain, and we indeed find that both boundaries coincide. Therefore, the EVM method provides a sharp boundary of the quantum domain for pure qubit test states, as one might expect from the results of Sec.~\ref{sec:pts}.

We recall that the mixing process inside the test-state source did not shift the boundary of the quantum domain for qubit states. In fact, this boundary can be reached by an optimized entanglement-breaking channel for any degree of mixing of the test states. The optimal POVM is that of minimum error discrimination. This may seem surprising, because the error probability, which depends on the fidelity (or the overlap) of the test states when they are pure, is a function of the trace distance between the test states when they are mixed. As it turns out, the well known relation between fidelity and trace distance
\begin{equation}
	D_{\rm tr}(\rho,\sigma) \le \sqrt{1 - F(\rho,\sigma)}
\end{equation}
is fulfilled with equality if $\rho$ and $\sigma$ are qubit states with equal degree of mixing. This can be shown by direct calculation with the help of the results in Ref.~\cite{chen02a}. Hence, for the investigated qubit system, mixed signal sources can be incorporated in the EVM-criterion without adding complexity to the problem. The resulting boundary of the quantum domain is tight, i.e., the quantum domain determined by the EVM-method reaches the boundary of the classical domain.\\

\emph{Squeezed thermal states.} In Sec.~\ref{sec:pts}, we derived the optimal measure and re-prepare strategy for an ensemble of pure test states. In part, we can use these results when considering mixed test states. For instance, it was argued that the re-prepared states $\{\ket{\tilde{\psi}_i}\}$ should be drawn from the set of squeezed vacuum states. Naturally, this is also true for mixed test states. The optimal POVM, however, is of a very different structure. While two pure test states span a two-dimensional space and the POVM elements are consequently represented by two-dimensional matrices, the squeezed thermal states act on an infinite-dimensional Hilbert space. Nevertheless, the optimal POVM can be found analytically.

We recall the above figure of merit for squeezed input states: The optimal entanglement-breaking channel minimizes ${\rm Var}_0(\hat p)$ for a fixed value of ${\rm Var}_0(\hat x)$. It is instructive to start with the simplified problem of finding the global minimum of ${\rm Var}_{\rho_0^{out}}(\hat p)$ while ignoring ${\rm Var}_{\rho_0^{out}}(\hat x)$ for the moment. The only constraints on the minimization are that $\rho_i^{out}$ are the results of an entanglement-breaking channel acting on the input states $\rho_i^{in}$, and Eqs.~(\ref{eq:symm1}) and (\ref{eq:symm2}) must be fulfilled. In fact these last two constraints are automatically satisfied if we consider entanglement-breaking strategies which obey a certain symmetry. If $\{ \pi_i, \ket{\tilde{\psi}_i} \}$ is a particular strategy which fulfills Eqs.~(\ref{eq:symm1}) and (\ref{eq:symm2}), then the phase-rotated strategy $\{ \hat{U}_{\frac{\pi}{2}} \pi_i \hat{U}_{\frac{\pi}{2}}^\dagger, \hat{U}_{\frac{\pi}{2}} \ket{\tilde{\psi}_i} \}$ leads to the same observations. Here,
\begin{equation}
	\hat{U}_{\frac{\pi}{2}} = \exp(-\bbi \frac{\pi}{2} \hat{n}).
\end{equation}
is a phase shift operator. An equal mixture of both strategies will then automatically satisfy the constraints (\ref{eq:symm1}) and (\ref{eq:symm2}). We find
\begin{widetext}
	\begin{align}\nonumber
		{\rm Var}_{\rho_0^{out}}(\hat p) & = \frac{1}{2} \left(\sum_i \tr(\pi_i \rho_0^{in}) \braket{\tilde{\psi}_i | \hat{p}^2 | \tilde{\psi}_i} + \sum_i \tr(\hat{U}_{\frac{\pi}{2}} \pi_i \hat{U}_{\frac{\pi}{2}}^\dagger \rho_0^{in}) \braket{\tilde{\psi}_i | \hat{U}_{\frac{\pi}{2}}^\dagger \hat{p}^2 \hat{U}_{\frac{\pi}{2}}| \tilde{\psi}_i} \right) \\ \nonumber
		& = \frac{1}{2} \left(\sum_i \tr(\pi_i \rho_0^{in}) \braket{\tilde{\psi}_i | \hat{p}^2 | \tilde{\psi}_i} + \sum_i \tr(\pi_i \rho_1^{in}) \braket{\tilde{\psi}_i | \hat{x}^2 | \tilde{\psi}_i} \right) \\ \label{eq:varp}
		& = \frac{1}{4}\left( \sum_i \tr(\pi_i \rho_0^{in}) \bbe^{-2 r_i} + \sum_i \tr(\pi_i \rho_1^{in}) \bbe^{2 r_i} \right).
	\end{align}
\end{widetext}
For each term in the summation, we can find the optimal degree of squeezing, $r_i$, and insert it into the above expression. We find
\begin{equation}\label{eq:varx}
	{\rm Var}_{\rho_0^{out}}(\hat p) = \frac{1}{2} \sum_i \sqrt{\tr(\pi_i \rho_0^{in}) \tr(\pi_i \rho_1^{in})}.
\end{equation}
For the remaining minimization over the POVM elements, we borrow a result from Ref.~\cite{fuchs95a}, namely,
\begin{equation}\label{eq:minfid}
	\min_{\pi_i} \sum_i \sqrt{\tr(\pi_i \rho_0^{in}) \tr(\pi_i \rho_1^{in})}  = F(\rho_0^{in}, \rho_1^{in}),
\end{equation}
where $F$ denotes the Uhlmann fidelity defined in Eq.~\ref{eq:fid}.

In summary, the smallest variance achievable by a measure and re-prepare strategy is given by half the fidelity of the two test states (see Fig.~\ref{fig:unitaries}). This is in agreement with the results of the entanglement verification. The boundary of the quantum domain is therefore tight at this minimum point. But is it tight elsewhere, i.e., for different values of ${\rm Var}_{\rho_0^{out}}(\hat x)$?

\begin{figure}
	\centering
		\includegraphics[width=0.4\textwidth]{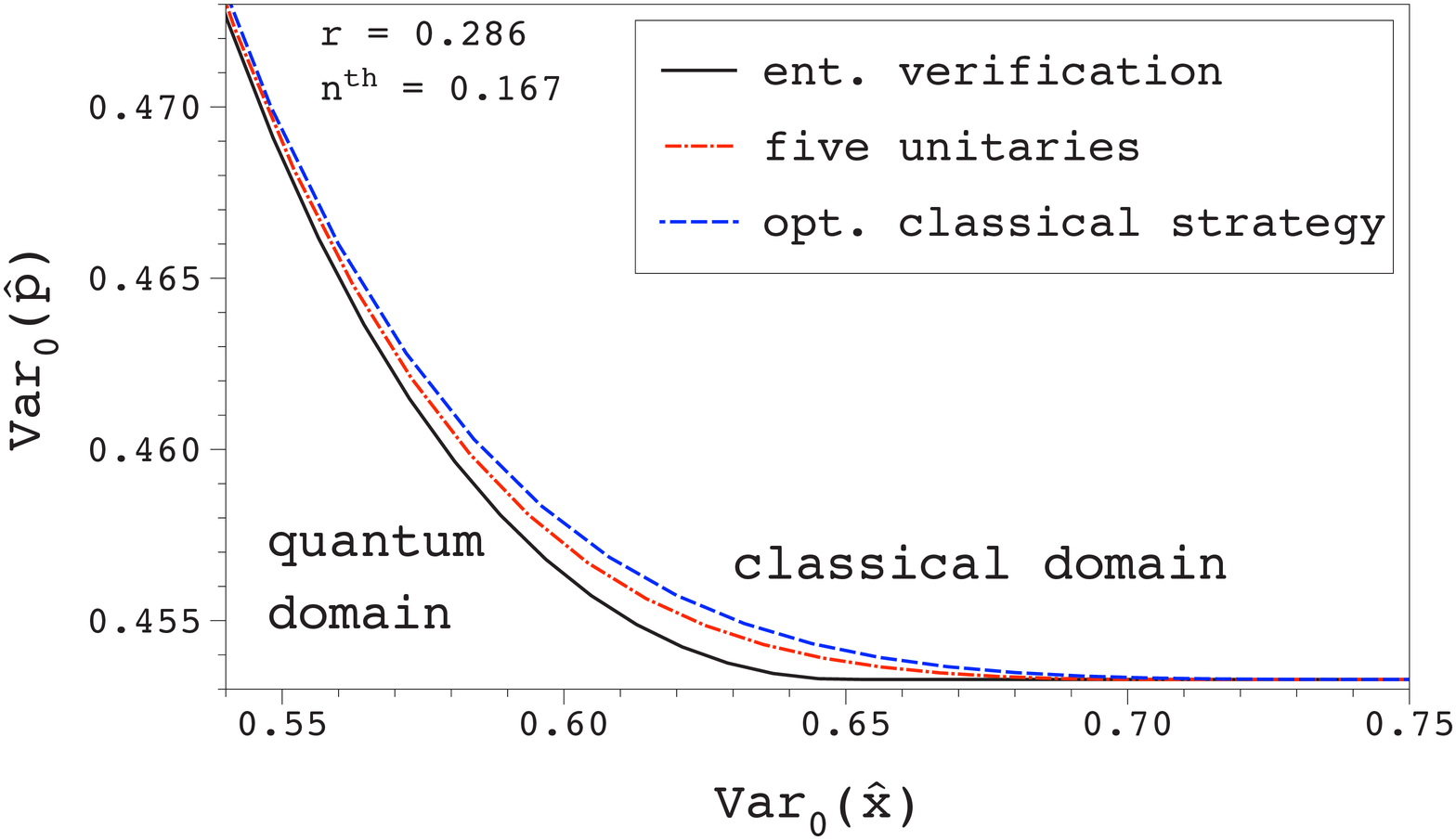}
	\caption{(Color online) Optimal measure and re-prepare strategy for squeezed thermal test states. Solid: entanglement verification with no information on the purifying system; dashed-dotted: five additional unitary operators on system $C$ in the EVM; dashed: optimized measure and re-prepare strategy.}
	\label{fig:mixattack}
\end{figure}

To answer this question, we introduce the additional constraint
\begin{equation}\label{eq:con}
	{\rm Var}_{\rho_0^{out}}(\hat x) = c,
\end{equation}
to the optimization problem, and a plot like Fig.~\ref{fig:mixattack} is obtained by varying $c$ in steps and minimizing ${\rm Var}_{\rho_0^{out}}(\hat p)$ at each step. This problem can be solved using Lagrange multipliers, i.e., we can minimize the Lagrange function
\begin{equation}
	\mathcal{L} = {\rm Var}_{\rho_0^{out}}(\hat p) + \lambda ({\rm Var}_{\rho_0^{out}}(\hat x) - c),
\end{equation}
which is now unconstrained, but the minimization runs over $\{ \pi_i \}$, $\{r_i\}$, and $\lambda$. Since both variances are linear in $\rho_0^{in}$ and in $\rho_1^{in}$, we can write the above expression as
\begin{equation*}
	\begin{split}
		\mathcal{L} = \frac{1}{4} \Bigl( \sum_i \bbe^{-2r_i} \tr(\pi_i [\rho_0^{in} + \lambda \rho_1^{in}]) \\
		+ \bbe^{2r_i} \tr(\pi_i [\rho_1^{in} + \lambda \rho_0^{in}]) \Bigr) - \lambda c.
	\end{split}
\end{equation*}
Defining the new states $\tilde{\rho}_0 = (\rho_0^{in} + \lambda \rho_1^{in})/(1+\lambda)$ and $\tilde{\rho}_1 = (\rho_1^{in} + \lambda \rho_0^{in}) /(1+\lambda)$, and minimizing over the $r_i$, we arrive at
\begin{align}
	\mathcal{L} & = \frac{1+\lambda}{2} \sum_i \sqrt{\tr(\pi_i \tilde{\rho}_0) \tr(\pi_i \tilde{\rho}_1)} - \lambda c\\
	\min_{\pi_i} \mathcal{L} & = \frac{1+\lambda}{2} F(\tilde{\rho}_0,\tilde{\rho}_1) - \lambda c.
\end{align}
In the last step, the relation (\ref{eq:minfid}) was used again. Now we can minimize this function with respect to $\lambda$, subject to the original constraint (\ref{eq:con}). Although this must still be done numerically because of the fidelity function, the problem is considerably simpler than a complete optimization over the POVM elements.

As it turns out, this method does not work for all values of $c$. Drawing the analogy to the pure-state case and Fig.~\ref{fig:pure}, only the optimal strategy in region (A) can be explained thus. In region (B), the Lagrange parameter $\lambda$ becomes negative, and the states $\tilde{\rho}_0$ and $\tilde{\rho}_1$ turn unphysical. For pure test states, the optimal measurement in region (B) depended on unambiguous state discrimination. Such a measurement strategy is not directly applicable \cite{raynal03a} to two squeezed thermal states, since they have equal support. It is still possible to construct a measure and re-prepare strategy which reaches the straight line bounding the quantum domain. A specific strategy which approximates the optimum arbitrarily well is presented in Appendix \ref{sec:appA}.

We now see that the entanglement verification method does not detect all states in the quantum domain, but increasing the size of $\chi(\rho_{ABC})$ and including more information about system $C$ brings a notable improvement.

\section{Conclusion}\label{sec:d}
In conclusion, a generic method to test quantum channels was presented. The method requires very few experimental resources and is therefore directly applicable to quantum memory or teleportation experiments. To accommodate tests with mixed states, we extended the concept of effective entanglement. Verification of such effective entanglement was investigated using the expectation value matrix in conjunction with partial transposition. A strong, yet simple extension of the expectation value matrix method for pure-state sources was proposed, where a source of mixed test states is approximated by a source of purifications of these test states. This approximation could detect all entanglement for the considered qubit state example. For two continuous-variable settings, where displaced thermal states and squeezed thermal states were considered, using purifications of the true test states lead to strong, but sub-optimal criteria. A procedure to improve the EVM method for these cases was proposed and conjectured to lead to optimal entanglement criteria. Furthermore, optimized measure and re-prepare strategies were considered as the counterparts of entanglement-preserving channels, and as a means of probing the strengths of the entanglement criteria.

The results obtained for modeled channels suggest that the derived benchmarks are reachable by current experiments.

\section{Acknowledgments}
We would like to thank Tobias Moroder for helpful discussion, Alex Lvovsky and Mirko Lobino for their input on the experimental feasibility, and Sarah Croke for her help on the optimized POVMs. This work was funded by the EU integrated project QAP, the Canadian NSERC Discovery Grant, and by Quantum Works.

\appendix

\section{USD-Type Strategy for Mixed States}\label{sec:appA}
\begin{figure}
	\centering
		\includegraphics[width=0.42\textwidth]{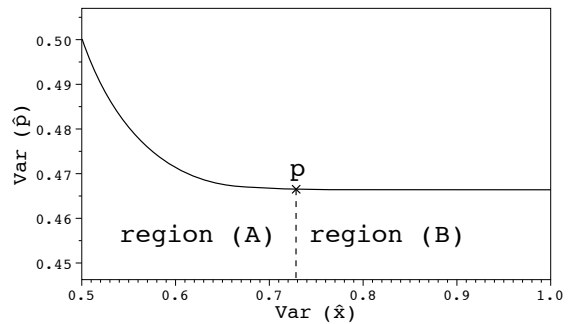}
	\caption{Schematic plot of the two different optimal measure and re-prepare strategies.}
	\label{fig:schematic}
\end{figure}

Here, we derive a measure and re-prepare strategy for two squeezed thermal states as test states, for the region (B) (Fig.~\ref{fig:schematic}). Again, we use the fact that the measured variances are linear functions of $\rho_0^{out}$, so that any convex combination of two measure and re-prepare strategies results in another valid entanglement-breaking channel. It therefore suffices to consider the slope of the straight line which joins the point $p$ in Fig.~\ref{fig:schematic} and a the point arising from an entanglement-breaking channel in the limit ${\rm Var}_{\rho_0^{out}} (\hat x) \rightarrow \infty$.

Let the coordinates of $p$ be denoted by $(v_x,v_p)$, then the slope of the straight line joining $p$ and any other point is given by
\begin{equation}\label{eqn:m}
	m = \frac{{\rm Var}_{\rho_0^{out}} (\hat p) - v_p}{{\rm Var}_{\rho_0^{out}} (\hat x) - v_x},
\end{equation}
where ${\rm Var}_{\rho_0^{out}} (\hat p)$ is given by Eq.~(\ref{eq:varp}) and the variance of $\hat x$ is defined accordingly.

The strategy which minimizes this slope is essentially an approximation to unambiguous state discrimination. We define a three-element POVM with $\hat{\Pi}_0$, $\hat{\Pi}_1 = \hat{U}_{\pi/2}^\dagger \hat{\Pi}_0 \hat{U}_{\pi/2}$, and $\hat{\Pi}_?$, and corresponding re-prepared states $\ket{\tilde{\psi}_0} = \ket{r}$, $\ket{\tilde{\psi}_1} = \ket{-r}$, and $\ket{\tilde{\psi}_?} = \ket{0}$. Consequently,
\begin{align*}
	{\rm Var}_{\rho_0^{out}} (\hat p) = \frac{1}{2}(\tr(\rho_0^{in} \pi_0)\bbe^{-2 r} + \tr(\rho_1^{in} \pi_0)\bbe^{2 r} + \tr(\rho_0^{in} \pi_?)) \\
	{\rm Var}_{\rho_0^{out}} (\hat x) = \frac{1}{2}(\tr(\rho_0^{in} \pi_0)\bbe^{2 r} + \tr(\rho_1^{in} \pi_0)\bbe^{-2 r} + \tr(\rho_0^{in} \pi_?)).
\end{align*}
Inserting into Eq.~(\ref{eqn:m}) and taking the limit $r \rightarrow \infty$, we find
\begin{equation}
	\lim_{r \to \infty} m = \frac{\tr(\rho_1^{in} \pi_0)}{\tr(\rho_0^{in} \pi_0)}.
\end{equation}
All that is left to do now is to find a POVM element $\pi_0$ for which the slope $m$ tends to zero. Setting $\pi_0 = \ket{\alpha}\bra{\alpha}$, where $\ket{\alpha}$ is a coherent state, achieves this. In this case, $m$ becomes a quotient of Husimi $Q$-functions:
\begin{equation}\label{eq:mq}
	m = \frac{\braket{\alpha |\rho_1| \alpha}}{\braket{\alpha |\rho_0| \alpha}} = \frac{Q_1(\alpha)}{Q_0(\alpha)}.
\end{equation}
The $Q$-function $Q_i(\alpha)$ of the squeezed thermal state $\rho_i$, $i \in \{0,1\}$, is given by \cite{marian93a}
\begin{equation*}
	\frac{1}{\pi\sqrt{(1+A_i)^2 - |B_i|^2}} \exp[(\tilde{A}_i-1) |\alpha|^2 - \frac{1}{2}(\tilde{B}_i \alpha^{*2} + \tilde{B}_i^* \alpha^2)],
\end{equation*}
with the definitions
\begin{align}
	A_0 & = \bar{n} + (2\bar{n} + 1) \sinh^2(r) = A_1 \\
	B_0 & = -(2\bar{n} + 1) \sinh(r) \cosh(r) = - B_1 \\
	\tilde{A}_0 & = \frac{\bar{n}(\bar{n}+1)}{\bar{n}^2 + (\bar{n} + \frac{1}{2})(1 + \cosh(2r))} = \tilde{A}_1 \\
	\tilde{B}_0 & = -\frac{(\bar{n} + \frac{1}{2}) \sinh(2r)}{\bar{n}^2 + (\bar{n} + \frac{1}{2})(1 + \cosh(2r))} = -\tilde{B}_1.
\end{align}
Inserting these expressions into Eq.~(\ref{eq:mq}), we find
\begin{equation*}
	m = \frac{\exp[- \frac{1}{2} \tilde{B}_1 (\alpha^{*2} + \alpha^2)]}{\exp[\frac{1}{2} \tilde{B}_1 (\alpha^{*2} + \alpha^2)]} = \exp[- \tilde{B}_1 (\alpha^{*2} + \alpha^2)].
\end{equation*}
For increasing $\alpha \in \mathbbm{R}$, this final expression tends to zero exponentially fast. Therefore, it is possible to reach the boundary of the quantum domain in Fig.~\ref{fig:schematic} for the whole parameter domain.

\end{document}